\documentclass[12pt]{article}
\usepackage{latexsym}
\begin{document}

\title{A local gauge invariant infrared regularization of the Yang-Mills theory.}
\author{A.A.Slavnov \\ Steklov Mathematical Institute\\ and Moscow State University \\ Moscow}

\maketitle

\begin{abstract}
A local gauge invariant infrared regularization for the Yang-Mills
theory is constructed on the basis of a higher derivative
formulation of the model.
\end{abstract}

\section{Introduction.}

In spite of great achievments in the study of nonabelian gauge
theories the problem of construction of a gauge invariant infrared
regularization remained unsolved. Of course in the Yang-Mills theory
a scattering matrix connecting the free asymptotic states, which
include massless particles, does not exist.  However some gauge
invariant infrared regularization allowing to make sense of formal
manipulations with it certainly would be welcome. Moreover in
nondiagonal gauges even the Green functions are plagued with
infrared divergencies which have to be regularized in a gauge
invariant way.

In this paper I propose an infrared regularization for the
Yang-Mills theory, which may be described by a local gauge invariant
Lagrangian. This Lagrangian contains higher derivatives and hence
the regularized theory includes nonpositive norm states, however in
the limit when the regularization is removed the nonphysical states
decouple.

In the next section a path integral formulation of such a
regularization will be presented.  The third section deals with the
field theoretical realization of this construction.

\section{A path integral regularization.}

To illustrate the main idea I present a heuristic derivation of the
regularized action for the $SU(2)$ gauge theory. Generalization to
other groups is straight forward.

The following formal equality obviously holds:
\begin{eqnarray}
 \int \exp\lbrace i \int [L_{YM}+m^2\varphi^*\varphi]dx\rbrace d\mu= \nonumber\\
\int \exp\lbrace i \int
[L_{YM}+m^{-2}(D^2\varphi')^*(D^2\varphi')-d^*D^2b-b^*D^2d]dx\rbrace
d\mu' \label{1}
\end{eqnarray}
Here $L_{YM}$ is the usual Yang-Mills Lagrangian
\begin{equation}
 L_{YM}=- \frac{1}{4}F_{\mu\nu}^jF_{\mu\nu}^j
\label{2}
\end{equation}
and the complex scalar fields $\varphi$ form the $SU(2)$ doublet,
which may be conveniently parameterised as follows:
\begin{equation}
 \varphi_1= \frac{iB^1+B^2}{\sqrt{2}}; \quad \varphi_2= \frac{B^0+iB^3}{\sqrt{2}}
\label{2a}
\end{equation}
Anticommuting complex scalar fields $b,d$ form the similar doublets.
The measure $d\mu$ includes differentials of all fields as well as
gauge fixing factors. The measure $d\mu'$ differs by the presence of
the differentials of the fields $b,b^*,d,d^*$. The operator $D^2$
denotes the sum $\sum_{\mu}D_{\mu}D_{\mu}$, where $D_{\mu}$ is the
covariant derivative
\begin{equation}
 D_{\mu} \varphi=(\partial_{\mu}+ \frac{ig\tau^j}{2}A_{\mu}^j)\varphi
\label{3}
\end{equation}
The boundary conditions for the Yang-Mills field $A_{\mu}$ are the
standard ones. The fields $\varphi$ are fastly decreasing, and the
fields $\varphi',b,d$ satisfy the Feynman boundary conditions. The
integral over $\varphi$  obviously produces a trivial constant, so
that l.h.s. of the eq.(\ref{1}) is just the path integral for the
Yang-Mills theory. Performing explicitely the integration over
$\varphi', b, d$ in the r.h.s. of the eq.(\ref{1}), we get the same
result.

The equation (\ref{1}) is formal, as neither l.h.s. nor r.h.s. exist
because of infrared divergencies. We define the infrared regularized
theory in the following way. Let us add to the action in the r.h.s.
the gauge invariant term
\begin{equation}
 \int \lbrace \alpha (D_{\mu}\varphi')^*(D_{\mu}\varphi')- \alpha m^2(d^*b+b^*d)\rbrace dx
\label{4}
\end{equation}
The integral in the r.h.s. of eq.(\ref{1}) is still infrared
divergent. However if we make the shift
\begin{equation}
 \varphi'\rightarrow \varphi'+ \hat{a}, \quad \hat{a}_1=0, \quad  \hat{a}_2=a,
\label{5}
\end{equation}
the regularized action acquires a form
\begin{eqnarray}
A_R= \int \lbrace - \frac{1}{4}F_{\mu \nu}^jF_{\mu \nu}^j-m^{-2}(D^2 \varphi)^*(D^2 \varphi)-(D_{\mu}d)^*D_{\mu}b-(D_{\mu}b)^*D_{\mu}d\nonumber\\
-\frac{a^2g^2}{4m^2}(\partial_{\mu}A_{\mu})^2+\frac{ag}{\sqrt{2}m^2}\partial^2B^j
\partial_{\mu}A_{\mu}^j+\nonumber\\
\alpha(D_{\mu}\varphi)^*(D_{\mu}\varphi)+ \frac{\alpha
g^2a^2}{4}A_{\mu}^2 - \frac{\alpha ga}{2 \sqrt{2}}B^j
\partial_{\mu}A_{\mu}^j- \alpha m^2(d^*b+b^*d)+ \ldots \label{6}
\end{eqnarray}
Here $\ldots$ denote the interaction terms which arise due to shift
(\ref{5}).

One sees that the shift (\ref{5}) generates the mass term for the
vector field, the term $(\partial_{\mu}A_{\mu})^2$, the mixing of
$B^j$ with $\partial_{\mu}A_{\mu}^j$ and additional interaction
terms. For simplicity in the following we choose $g^2a^2=2m^2$. Then
the mass of the Yang-Mills field is $\sqrt{\alpha}m$.

The theory described by the action (\ref{6}) is free of infrared
singularities. At the same time the action is local and invariant
with respect to the gauge transformations
\begin{eqnarray}
 A_{\mu}^j \rightarrow A_{\mu}^j -g\varepsilon^{jik}A_{\mu}^i\eta^k+ \partial_{\mu}\eta^j\nonumber\\
B^0 \rightarrow B^0+g(B^j \eta^j)\nonumber\\
B^j \rightarrow B^j-m \eta^j- \frac{g}{2} \varepsilon^{jik}B^i
\eta^k- \frac{g}{2}B^0 \eta^j \label{7}
\end{eqnarray}
This invariance allows to use in the corresponding path integral any
admissible gauge condition. Particularly convenient is the Lorentz
gauge $\partial_{\mu}A_{\mu}=0$. In this gauge the mixing between
$A_{\mu}^j$ and $B^j$ is absent and renormalizability is manifest.

One has to understand that the transformation (\ref{5}) is not a
simple change of variables. It changes the boundary conditions in
the path integral. Rather it is a definition of the infrared
regularized Yang-Mills theory. More precisely
\begin{equation}
 \int \exp \lbrace i \int L_{YM}dx \rbrace d \mu|_{reg}= \int \exp \lbrace iA_R \rbrace d
 \mu'
\label{8}
\end{equation}

The equation (\ref{8}) gives a definition of the infrared
regularized scattering matrix for the Yang-Mills theory as a path
integral of the exponent of a local gauge invariant action. It also
allows to give a sensible definition of the correlation functions as
in the regularised theory one can perform the Wick rotation in all
Feynman integrals making the transition $\alpha \rightarrow 0$
legitimate.

In the next section we shall show that this path integral
regularization admits an elegant field theoretical realization,
similar to the BRST quantization of gauge invariant models.

\section{Canonical quantization and unitarity of regularized theory.}

It was shown in our papers (\cite{Sl1}, \cite{Sl2}) that a change of
variables in a path integral which introduces higher derivatives may
be interpreted as a transition to a field theory model including
unphysical ghost fields. This theory posesses a (super)symmetry
which leads via Noether theorem to existence of a conserved
nilpotent charge $Q$. Existence of such a charge allows to separate
the physical states by imposing the condition
\begin{equation}
 Q|\psi>_{phys}=0
\label{12}
\end{equation}
These states have nonnegative norms and the scattering matrix is
unitary in the subspace (\ref{12}).

Below we shall show that a similar construction may be done in the
present model. A peculiar feathure of our model is related to the
fact that contrary to the cases considered before the conserved
charge $Q$ is not nilpotent. Nilpotency is recovered only in the
limit $ \alpha \rightarrow 0$, and this limit, when it exists,
determines the Yang-Mills theory. The limit $ \alpha \rightarrow 0$
for the on-shell $S$-matrix does not exist due to infrared
divergencies, but the formal expression for the matrix elements in
the limit when the regularization is removed coincides with the
$S$-matrix elements of original Yang-Mills theory.

Our starting point is the regularized action
\begin{eqnarray}
 A_R= \int \{ - \frac{1}{4} F_{\mu \nu}^jF_{\mu \nu}^j-m^{-2}(D^2(\varphi+ \hat{a}))^*D^2(\varphi+ \hat{a})+(D_{\mu}d)^*D_{\mu}b\nonumber\\
+(D_{\mu}b)^*D_{\mu}d + \alpha[(D_{\mu}(\varphi+
\hat{a}))^*D_{\mu}(\varphi+ \hat{a})-m^2(d^*b+b^*d)]\}dx \label{13}
\end{eqnarray}
This action is invariant with respect to the gauge transformations
(\ref{7}) and the supersymmetry transformations
\begin{eqnarray}
 \delta \varphi= \varepsilon b\nonumber\\
 \delta d= m^{-2}D^2(\varphi+ \hat{a})\varepsilon
\label{14}
\end{eqnarray}
where $\varepsilon$ is an anti-Hermitean parameter anticommuting
with $b,d$. In terms of the components this transformation looks as
follows:
\begin{eqnarray}
 \delta \varphi= \varepsilon b\nonumber\\
\delta d_1=[m^{-2}(D^2 \varphi)_1 + \frac{1}{\sqrt{2}m}(i \partial_{\mu}A_{\mu}^1+ \partial_{\mu}A_{\mu}^2)] \varepsilon\nonumber\\
\delta d_2=[m^{-2}(D^2 \varphi)_2 -
\frac{i}{\sqrt{2}m}(\partial_{\mu}A_{\mu}^3- \frac{g}{2m}A_{\mu}^2]
\varepsilon \label{15}
\end{eqnarray}
Note that these transformations are not nilpotent: $ \delta^2d \neq
0$. The nilpotency is restored only in the limit $\alpha=0$.

The action (\ref{13}) is invariant both with respect to the gauge
transformations and the transformations (\ref{15}). The
supersymmetry transformations do not change the fields $A_{\mu}$, so
it is convenient to choose for quantization a manifestly
supersymmetric and renormalizable gauge $\partial_{\mu}A_{\mu}=0$.

In this gauge the Lagrangian may be written in terms of the
components $B^a, B^0$, and the similar components for the fields $b,
d$
\begin{eqnarray}
 b_1= \frac{ib^1+b^2}{\sqrt{2}}; \quad b_2= \frac{b^0+ib^3}{\sqrt{2}}\nonumber\\
d_1= \frac{d^1-id^2}{\sqrt{2}}; \quad d_2=
\frac{-id^0+d^3}{\sqrt{2}} \label{16}
\end{eqnarray}
as follows
\begin{eqnarray}
 L_R=- \frac{m^{-2}}{2} \partial^2 B^{\rho} \partial^2 B^{\rho}-i \partial_{\mu}b^{\rho} \partial_{\mu}d^{\rho}- \frac{1}{4}(\partial_{\mu}A_{\nu}^a- \partial_{\nu}A_{\mu}^a)^2\nonumber\\
+ \alpha[\partial_{\mu}B^{\rho}
\partial_{\mu}B^{\rho}-im^2b^{\rho}d^{\rho}]+ \ldots \label{17}
\end{eqnarray}
where $\rho=0,1,2,3$ and $\ldots$ denote the interaction terms.

The quantization of the Yang-Mills fields $A_{\mu}^j$ is performed
in a standard way and requires the introduction of the Faddeev-Popov
ghosts $\bar{c},c$. The scalar fields $b,d$ also make no problems.

The fields $B^{\rho}$ are described by the higher derivative
Lagrangian and for their quantization we shall use Ostrogradsky
canonical formalism (\cite{Sl3}, \cite{Sl1}, \cite{Sl2}). As we are
working in the framework of perturbation theory it is sufficient to
consider the quantization of the free theory.

In the Ostrogradsky formalism the system is described by the
canonical coordinates
\begin{equation}
 X_1^{\rho}=B^{\rho}; \quad X_2^{\rho}= \dot{B}^{\rho}
\label{18}
\end{equation}
and conjugate momenta
\begin{eqnarray}
 P_1^{\rho}= \frac{\delta L}{\delta \dot{B}^{\rho}}- \partial_0(\frac{\delta L}{\delta \ddot{B}^{\rho}})= \alpha \partial_0B^{\rho}+m^{-2} \partial_0 \partial^2 B^{\rho}\nonumber\\
P_2^{\rho}= \frac{\delta L}{\delta \ddot{B}^{\rho}}=-m^{-2}
\partial^2 B^{\rho} \label{19}
\end{eqnarray}
The Hamiltonian for the $B^{\rho}$ fields, being written in terms of
Fourier components is given by the expression
\begin{eqnarray}
H=P_1^{\rho}X_2^{\rho}+ P_2^{\rho} \dot{X}_2^{\rho}-L=\nonumber\\
P_1^{\rho}X_2^{\rho}-
\frac{m^2}{2}(P_2^{\rho})^2-k^2P_2^{\rho}X_1^{\rho}-
\frac{\alpha}{2}(X_2^{\rho})^2+ \frac{\alpha
k^2}{2}(X_1^{\rho})^2+\ldots \label{20}
\end{eqnarray}
Introducing the creation and annihilation operators one can write
the free hamiltonian in the form
\begin{equation}
 H^0= \int [\omega_1(k)q_1^{\rho+}(k)q_1^{\rho-}(k)-\omega_2(k)q_2^{\rho+}(k)q_2^{\rho-}(k)]dk
\label{21}
\end{equation}
where
\begin{equation}
 q_1^{\rho\pm}= \frac{\pm i \omega_1 \alpha X_1^{\rho}+P_1^{\rho} \mp i\omega_1P_2^{\rho}}{\sqrt{2 \alpha \omega_1}}; \quad q_2^{\rho \pm}= \frac{-\alpha X_2^{\rho} \mp i \omega_2P_2^{\rho}+P_1}{\sqrt{2\alpha \omega_2}}
\label{22}
\end{equation}
In these equations $\omega_1= \sqrt{k^2}; \quad \omega_2= \sqrt{k^2+
\alpha m^2}$, and the operators $q_{1,2} ^{\rho \pm}$ satisfy the
commutation relations
\begin{equation}
 [q_1^{\rho-}(k),q_1^{\sigma+}(k')]= \delta^{\rho \sigma} \delta(k-k'): \quad [q_2^{\rho-}(k),q_2^{\sigma+}(k')]=- \delta^{\rho \sigma} \delta(k-k')
\label{23}
\end{equation}
One sees that the operators $q_2^{\rho+}$ create negative norm
states.

The free Hamiltonian for the supersymmetry ghosts
$b^{\rho},d^{\rho}$ may be obtained in a standard way
\begin{equation}
 H_0'=i \int \omega_2[d^{\rho+}(k)b^{\rho-}(k)-b^{\rho+}(k)d^{\rho-}(k)]dk
\label{24}
\end{equation}
where the creation and annihilation operators are given by the
equations
\begin{equation}
 d^{\rho \pm}= \frac{d^{\rho}\omega_2 \pm p_b^{\rho}}{\sqrt{2\omega_2}}; \quad b^{\rho \pm}= \frac{b^{\rho} \omega_2 \mp p_d^{\rho}}{\sqrt{2 \omega_2}}
\end{equation}
They satisfy the anticommutation relations
\begin{equation}
 [b^{\rho-}(k),d^{\sigma+}(k')]_+=-i \delta^{\rho \sigma} \delta(k-k'); \quad [d^{\rho-}(k),b^{\sigma+}(k')]_+=i \delta^{\rho \sigma} \delta(k-k')
\label{26}
\end{equation}

The space of states includes many unphysical exitations, like the
supersymmetry ghost states, states corresponding to the fields
$B^{\rho}$, unphysical components of $A_{\mu}$ and Faddeev-Popov
ghosts. The real physical states including only transversal
components of the Yang-Mills field may be separated by imposing on
the asymptotic states the conditions
\begin{equation}
 Q_0|\psi>_{phys}=0
\label{27}
\end{equation}
\begin{equation}
Q_0^{BRST}|\psi>_{phys}=0 \label{28}
\end{equation}
and taking the limit $\alpha \rightarrow 0$. Here $Q_0$ is the free
charge assosiated with the supersymmetry transformations(\ref{15}),
and $Q_0^{BRST}$ is the free BRST charge.

The invariance of the action (\ref{13}) with respect to the
supersymmetry transformations (\ref{15}) generates via Noether
theorem the conserved current, whose asymptotic form is
\begin{equation}
 J_{\mu}=m^{-2}(\partial_{\mu}  B^{\rho}b^{\rho}-  B^{\rho} \partial_{\mu}b^{\rho})
\label{29}
\end{equation}
The corresponding conserved charge may be written as follows
\begin{eqnarray}
 Q_0= \frac{1}{\sqrt{2 \omega_2}} \int \lbrace b^{\rho+}(P_1^{\rho}+i \omega_2P_2^ {\rho}- \alpha X_2^{\rho})+(P_1^{\rho}-i \omega_2P_2^{\rho}- \alpha X_2^{\rho})b^{\rho-}\rbrace dk\nonumber\\
\sim const \int \lbrace b^{\rho+}(k)
\frac{q_1^{\rho-}(k)+q_2^{\rho-}(k)}{2}+
\frac{q_1^{\rho+}(k)+q_2^{\rho+}(k)}{2}b^{\rho-}(k)\rbrace
dk+O(\alpha) \label{30}
\end{eqnarray}
One sees that although for a finite $ \alpha$ the charge $Q_0$ is
not nilpotent, in the limit $\alpha \rightarrow 0$ the nilpotency is
recovered as the operators $b^{\rho+}, b^{\rho-}$ and  $q_+^{\rho
\pm}=q_1^{\rho \pm}+q_2^{\rho \pm}$ are mutually (anti)commuting.

Any vector annihilated by $Q_0$ may be presented in the form
\begin{equation}
 |\varphi>=|\varphi>_A+Q_0|\chi>+O(\alpha)
\label{31}
\end{equation}
Here $|\varphi>_A$ is a vector which does not include the
exitations, corresponding to the ghost fields $q_{1,2}^{\rho}$ and
$b^{\rho}, c^{\rho}$. This vector depends only on the Yang-Mills
field exitations and the Faddeev-Popov ghosts. Imposing on it the
condition (\ref{28}), which is compatible with the condition
(\ref{27}), we conclude that the vectors $|\psi>_{phys}$ have a form
\begin{equation}
 |\psi>_{phys}=|\psi>_{tr}+|N>+O(\alpha)
\label{32}
\end{equation}
where $|\psi>_{tr}$ depends only on transversal polarizations of the
Yang-Mills field, and $|N>$ is a zero norm vector. Hence in the
limit $\alpha \rightarrow 0$ we recover the usual Yang-Mills theory.
It completes the proof.

\section{Discussion.}

In the present paper we proposed a local gauge invariant infrared
regularization of the Yang-Mills theory. Our construction is based
on the mechanism different from the mechanism commonly used in the
process of regularization. Usually one introduces in a regularized
theory some unphysical exitations which disappear from the spectrum
when the regularization is removed. In our scheme unphysical
exitations do not disappear when the regularization is removed , but
decouple completely from the physical exitations, which is
sufficient for a physical interpretation of the theory.

{\bf Acknowledgements} \\
This work was supported  by RBRF, under grant 050100541 and by the
RAS program "Theoretical problems of mathematics". The work was
completed when the author was visiting the University of Trento. I
thank I.Lazzizzera for hospitality and the Trento University
administration for support.


\begin{thebibliography}{99}{\small
\bibitem{Sl1} A.A.Slavnov, Phys.Lett.B, 258 (1991), 391.
\bibitem{Sl2} A.A.Slavnov, Phys.Lett.B, 620 (2005), 97.
\bibitem{Sl3} A.A.Slavnov, Nucl.Phys.B, 31 (1971), 301.}
\end{thebibliography}
\end{document}